\documentclass[12pt]{article}  

 \usepackage{graphicx}

 \newcommand{\comma}{\;\; ,}
 \newcommand{\period}{\;\; .}
 \newcommand{\eq}{\; = \;}
 \newcommand{\sep}{\;\; , \;\;}
 \newcommand{\be}{\begin{equation}}
 \newcommand{\bd}{\begin{displaymath}}
 \newcommand{\ee}{\end{equation}}
 \newcommand{\ed}{\end{displaymath}}
 \newcommand{\ba}{\begin{eqnarray}}
 \newcommand{\ea}{\end{eqnarray}}
 \renewcommand{\i}{{\rm i}}
 \newcommand{\e}{{\rm e}}
 \newcommand{\Wb}{\overline{W}}
 \newcounter{storeeqn}

 \title{The challenge of the chiral Potts model}

 \author{R. J. Baxter\\
  {\small Centre for Mathematics and its Applications}\\ 
   {\small The Australian National University, Canberra} \\
   {\small ACT 0200, Australia}}

 \date{\protect \small 25 Oct 2005 }
 \begin{document}

 \maketitle

 \begin{abstract}
 The chiral Potts model continues to pose particular challenges
 in statistical mechanics: it is ``exactly solvable'' in the sense that 
 it satisfies the Yang-Baxter relation, but actually obtaining the 
 solution is not easy. Its free energy was calculated in 1988 and the 
 order parameter was conjectured in full generality a year later. 
 However, a derivation of that conjecture had to wait until 2005. Here 
 we discuss that derivation.
 \end{abstract}.

 \section{Introduction}

 In 1970 I was in England, where my wife and I stayed for five months 
 with my parents in Essex. It was largely holiday, as we were on our 
 way back to Australia after two years in Boston, where I had been 
 introduced to the six-vertex models and the Bethe ansatz by Elliott
 Lieb.

 However, I did visit Cyril Domb's group at King's College, London, 
 and it was there that I first interacted with Tony Guttmann, who was 
 also visiting the department: he was 
 an invaluable aid to navigating the labyrinthine corridors and 
 staircases that linked the department's quarters in Surrey Street 
 with the main part of the College. 

 Tony's natural enthusiasm for statistical mechanics must have been 
 infectious, for it was at this time that I realised that the 
 transfer matrices of the six-vertex model commuted - a vital first 
 step in the subsequent solution of the eight-vertex model.  

 This led to the solution of a number of other two-dimensional lattice 
 models. One that has proved particularly challenging is the chiral 
 Potts model. Here I wish to discuss some of the insights that led to 
 the recent derivation of its order parameters.

 The chiral Potts model is a two-dimensional classical lattice model 
 in statistical  mechanics, where spins live on sites of a lattice
 and each spin takes $N$ values
 $0,1, \ldots, N-1$,  and adjacent spins interact with Boltzmann 
 weight functions  $W, \overline{W}$. We consider only the case when 
 the model 
 is ``solvable'', by which we mean that  $W, \overline{W}$ satisfy the 
 star-triangle (``Yang-Baxter'') relations \cite{BPAuY88}. The free 
 energy of  the infinite lattice was first obtained in 1988 by using 
 the invariance properties of the free energy and its
 derivatives.\cite{RJB88} Then in 1990 the functional transfer matrix 
 relations of Bazhanov and Stroganov \cite{BazStrog90} were used to 
 calculate  the free energy more explicitly as a double 
 integral.\cite{BBP90, RJB90, RJB91} The model has a critical 
	temperature, below which the system exhibits ferromagnetic order.

 The next step was to calculate the order parameters ${\cal M}_1, 
 \ldots , {\cal M}_{N-1}$ (defined below). These depend on a 
 constant $k$ which decreases from one to zero as the temperature 
 increases from zero to criticality. In 1989 Albertini {\it et al} 
 \cite{AMPT89} made the elegant conjecture, based on the available 
 series expansions, that 
 \be \label{conj}
 {\cal M}_r \eq k^{r(N-r)/N^2} \; \; , \; \;  0 \leq r \leq N    
 \period \ee
 It might have been expected that a proof of such a simple
 formula would not have been long in coming, but in fact it proved to
 be a remarkably difficult problem. Order parameters (spontaneous 
 magnetizations) are notoriously  more difficult to calculate than 
 free energies. For the Ising model (to which the chiral 
 Potts model reduces when $N=2$), the free energy was calculated by 
 Onsager in 1944 \cite{Onsager44}, but it was five years later when 
 at a conference in
 Florence he announced his result for the spontaneous magnetization, 
 and not till 1952 that the first published proof was given by 
 Yang\cite{Yang52,Onsager71}.

 Similarly, the free energy of the eight-vertex model was calculated
 in 1971.\cite{Baxter71} The spontaneous magnetization and polarization 
 were conjectured in 1973 and 1974, respectively\cite{BarberBax73, 
 BaxKelland74}, but it was not till 1982
 that a proof of the first of these conjectures were 
 published\cite{book82}. A proof of the second had to wait until 
 1993\cite{JMN93}!

	By then three separate methods had been used. The Onsager-Yang
 calculation was based on the particular 
 free-fermion/spinor/pfaffian/Clifford algebra structure of the Ising 
 model\cite{MPW63}. As far as the auther is aware, this has never been 
 extended to the other models: it would be very significant if 
 it could be.

 The eight-vertex and subsequent hard-hexagon calculation was made using
 the corner transfer matrix method, which had been discovered in
 1976\cite{Baxter76}. This worked readily for the magnetization
 (a single-site correlation), but not for the polarization (a 
  single-edge correlation). This problem was remedied by the 
  ``broken rapidity line'' technique discovered by Jimbo 
 {\it et al} \cite{JMN93}.

 For all the two-dimensional solvable models, the Boltzmann weight 
 functions  $W, \overline{W}$ depend on
 parameters $p$ and $q$. These parameters are known as {\em rapidities}
 and are associated with lines (the dotted lines of Figure
 \ref{sqlattice}) that run through the midpoints of the edges of the 
 lattice. In general these are complex numbers, or sets of
 related complex numbers. In all of the models we have mentioned, with 
 the notable exception of the $N > 2$ chiral Potts model, these  
 parameters can be chosen so that $W, \overline{W}$ depend only on the
 {\em rapidity difference} (spectral parameter) $p - q$. 

 This property seems to be an essential element in the corner transfer
 matrix method: the star-triangle relation ensures that the
 corner transfer matrices factor, but the difference property is then
 needed to show that the factors commute with one another and are 
 exponentials in the rapidities. The difference property  is {\em not} 
 possessed by the $N > 3$ chiral Potts model and one is unable to 
 proceed. At first the author thought this would prove to be 
 merely a technical complication and embarked on a low-temperature
 numerical  calculation\cite{Baxter93} in the hope this would reveal 
 the kind of simplifications that happen with the other models. This 
 hope was not realised.


 I then looked at the technique of Jimbo {\it et al} and in 1998 
 applied it to the chiral Potts model. One could write down functional 
 relations satisfied by the generalized order parameter ratio function
 $G_{pq}(r)$, and for $N=2$
 these were sufficient (together with an assumed but very plausible 
 analyticity property) to solve the problem. However, for $N > 2$ there 
 was still a difficulty. Then $p$, $q$ are points on an algebraic 
 curve of genus $> 1$ and there is no obvious uniformizing substitution. 
 The functional relations themselves do not define $G_{pq}(r)$: one 
 needs some additional analyticity information, and that seems hard to 
 come by.


 The calculation of the free energy of the chiral Potts model
 \cite{RJB90, RJB91, RJB03}
 proceeds in two stages. First one considers a related  
 ``$\tau_2 (t_q)$'' model.\cite{RJB04} This is intimately 
 connected with the superintegrable case of the chiral Potts 
 model.\cite{RJB89} It is much simpler than the chiral Potts model
 in that its Boltzmann weights depend on the horizontal
 rapidity $q$ only via a single parameter $t_q$, and are linear in 
 $t_q$. Its row-to-row transfer matrix is the product 
 of two chiral Potts transfer matrices, one with horizontal 
 rapidity $q$, the other with a related rapidity $r = V R q$
 defined by eqn. (\ref{autos}) of section 2. 

 For a finite lattice, the partition function $Z$ of the 
 $\tau_2 (t_q)$ model is therefore a polynomial in $t_p$.
 The free energy is the logarithm of $Z^{1/M}$, where $M$ is the 
 number of sites of the lattice, evaluated in the thermodynamic 
 limit when the lattice becomes infinitely big. This limiting
 function of  course may have singularities in the complex $t_q$
 plane.  {\it A  priori}, one might expect it to have $N$ branch 
 cuts, each running though one of the $N$ roots of unity. 
 However, one can argue that in fact it only has one such cut. As 
 a result the free energy (i.e. the maximum eigenvalue of the
 transfer matrix) can be calculated by a Wiener-Hopf factorization.

 The second stage is to factor this free energy to obtain 
 that of the chiral  Potts model.

 It was not until 2004 that I realised that :

 (1) If one takes $p$, $q$ to be related by eqn. (\ref{spcase}) below,
 then $G_{pq}(r)$ can be expressed in terms of partition functions
 that involve $p, q$ only via the Boltzmann weights of the 
 $\tau_2 (t_{p'})$ model, with $p' = R^{-1} p$.

 (2) It is {\em not} necessary to obtain $G_{pq}(r)$ for arbitrary $p$ 
  and $q$. To verify the conjecture (\ref{conj}) it is sufficient to 
 obtain it under the restriction (\ref{spcase}).

 I indicate the working in the following sections: a fuller account is
 given in Ref. \cite{RJB05b}. The calculation of  $G_{pq}(r)$ for 
 general $p$, $q$ remains an unsolved problem: still interesting, but 
	not necessary for the derivation of the order parameters ${\cal M}_r$.


 \section{Chiral Potts model}

 We use the notation of \cite{BPAuY88, BBP90, RJB98}. Let $k, k'$
 be two real variables in the range $(0,1)$, satisfying 
 \be k^2 + {k'}^2 = 1  \period \ee
 Consider four parameters
 $x_p, y_p, \mu_p, t_p$ satisfying the relations
 \be \label{xymu} 
 k x_p^N = 1-k'/\mu_p^N 
 \sep k y_p^N = 1-k'\mu_p^N \sep t_p = x_p y_p \period \ee

 Let $p$ denote the set $\{x_p, y_p, \mu_p, t_p \}$. Similarly, let 
 $q$ denote the set  $\{x_q, y_q, \mu_q, t_q \}$.
 We call $p$ and $q$ ``rapidity'' variables. Each has one free
 parameter and is a point on an algebraic curve.

 Define Boltzmann weight functions $W_{pq}(n), \Wb _{pq}(n)$ by
 \addtocounter{equation}{1}
 \setcounter{storeeqn}{\value{equation}}
 \setcounter{equation}{0}
 \renewcommand{\theequation}{\arabic{storeeqn}\alph{equation}}
 \ba \label{WWba}
 W_{pq}(n) & = & (\mu_p/\mu_q)^n \prod_{j=1}^n \frac{y_q - \omega^j x_p}
 {y_p - \omega^j x_q} \comma  \\
 \label{WWbb}
 \Wb_{pq}(n) & = &  (\mu_p \mu_q)^n  \prod_{j=1}^n \frac{\omega x_p -   
 \omega^j x_q} {y_q - \omega^j y_p} \comma \ea
 where
 \bd \omega \eq \e^{2\pi \i/N} \period \ed
 They satisfy the periodicity conditions
 \bd 
 W_{pq}(n + N) = W_{pq}(n) \sep \Wb_{pq}(n + N) = \Wb_{pq}(n) 
 \period \ed

 \setcounter{equation}{\value{storeeqn}}
 \renewcommand{\theequation}{\arabic{equation}}



 \setlength{\unitlength}{1pt}
 \begin{figure}[hbt]

 \begin{picture}(420,260) (0,0)

 \multiput(30,15)(5,0){73}{.}
 \multiput(30,75)(5,0){32}{\bf .}
 \multiput(31,75)(5,0){32}{\bf .}
 \multiput(202,75)(5,0){35}{\bf .}
 \multiput(203,75)(5,0){35}{\bf .}
 \multiput(30,135)(5,0){73}{.}
 \multiput(30,195)(5,0){73}{.}
 \put (190,72) {\line(0,1) {8}}
 \put (200,72) {\line(0,1) {8}}
 \thicklines

 \put (69,72) {\large $< $}
 \put (70,72) {\large $< $}
 \put (71,72) {\large $< $}

 \put (308,12) {\large $< $}
 \put (309,12) {\large $< $}
 \put (310,12) {\large $< $}

 \put (308,72) {\large $< $}
 \put (309,72) {\large $< $}
 \put (310,72) {\large $< $}

 \put (308,132) {\large $< $}
 \put (309,132) {\large $< $}
 \put (310,132) {\large $< $}

 \put (308,192) {\large $< $}
 \put (309,192) {\large $< $}
 \put (310,192) {\large $< $}

 \put (42,230) {\large $\wedge$}
 \put (42,229) {\large $\wedge$}
 \put (42,228) {\large $\wedge$}

 \put (102,230) {\large $\wedge$}
 \put (102,229) {\large $\wedge$}
 \put (102,228) {\large $\wedge$}

 \put (162,230) {\large $\wedge$}
 \put (162,229) {\large $\wedge$}
 \put (162,228) {\large $\wedge$}

 \put (222,230) {\large $\wedge$}
 \put (222,229) {\large $\wedge$}
 \put (222,228) {\large $\wedge$}

 \put (282,230) {\large $\wedge$}
 \put (282,229) {\large $\wedge$}
 \put (282,228) {\large $\wedge$}

 \put (342,230) {\large $\wedge$}
 \put (342,229) {\large $\wedge$}
 \put (342,228) {\large $\wedge$}

 \thinlines


 \put (176,102) {{\Large \it a}}
 \put (320,60) {{\Large \it q}}
 \put (83,60) {{\Large \it p}}
 \put (380,-2) {{\Large \it h}}
 \put (380,118) {{\Large \it h}}
 \put (380,178) {{\Large \it h}}

 \put (195,105) {\circle{7}}

 \put (16,45) {\line(1,-1) {60}}
 \put (16,165) {\line(1,-1) {180}}
 \put (76,225) {\line(1,-1) {117}}
 \put (198,103) {\line(1,-1) {117}}
 \put (196,225) {\line(1,-1) {180}}
 \put (316,225) {\line(1,-1) {60}}
 \put (16,165) {\line(1,1) {60}}
 \put (16,45) {\line(1,1) {180}}
 \put (76,-15) {\line(1,1) {117}}
 \put (198,107) {\line(1,1) {118}}
 \put (196,-15) {\line(1,1) {180}}
 \put (316,-15) {\line(1,1) {60}}

 \put (75,105) {\circle*{7}}
 \put (315,105) {\circle*{7}}
 \put (75,-15) {\circle*{7}}
 \put (195,-15) {\circle*{7}}
 \put (315,-15) {\circle*{7}}

 \put (15,45) {\circle*{7}}
 \put (135,45) {\circle*{7}}
 \put (255,45) {\circle*{7}}
 \put (375,45) {\circle*{7}}

 \put (15,165) {\circle*{7}}
 \put (135,165) {\circle*{7}}
 \put (255,165) {\circle*{7}}
 \put (375,165) {\circle*{7}}

 \put (75,225) {\circle*{7}}
 \put (195,225) {\circle*{7}}
 \put (315,225) {\circle*{7}}

 \put (42,-40) {{\Large \it v}}
 \put (102,-40) {{\Large \it v}}
 \put (162,-40) {{\Large \it v}}
 \put (222,-40) {{\Large \it v}}
 \put (282,-40) {{\Large \it v}}
 \put (342,-40) {{\Large \it v}}

 \multiput(45,-25)(0,5){52}{.}
 \multiput(105,-25)(0,5){52}{.}
 \multiput(165,-25)(0,5){52}{.}
 \multiput(225,-25)(0,5){52}{.}
 \multiput(285,-25)(0,5){52}{.}
 \multiput(345,-25)(0,5){52}{.}
 \end{picture}

 \vspace{1.5cm}
 \caption{\footnotesize  The square lattice (solid lines, drawn 
 diagonally), and the associated rapidity lines (broken or dotted).}
 \label{sqlattice}
 \end{figure}
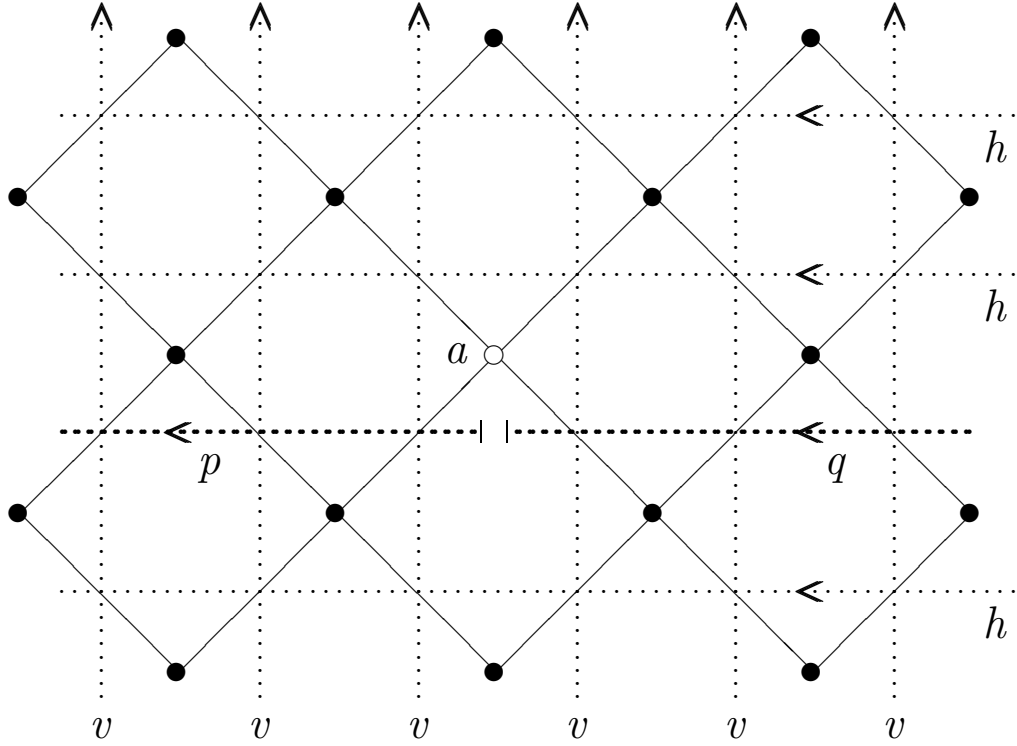

 Now consider the square lattice $\cal L$, drawn diagonally as in 
 Figure \ref{sqlattice}, with a total of $M$ sites. On each site 
 $i$ place a  spin $\sigma_i$, which can take any one of the $N$ 
 values $0, 1, \ldots, N-1$. 

 The solid lines in Figure \ref{sqlattice} are the edges of 
 $\cal L$. Through each such edge  there pass two dotted or 
 broken lines - a vertical line denoted $v$ and a horizontal line
 denoted $h$ (or $p$ or $q$). These $v, h, p, q$ are rapidity
 variables, as defined above. We refer to each dotted line as a 
 ``rapidity line''.

 With each SW - NE edge $(i,j)$ (with $i$ below $j$) associate an 
 edge weight  $W_{vh}(\sigma_i - \sigma_j)$. 
 Similarly, with each SW - NE edge $(j,k)$  ($j$ below $k$), associate
 an edge weight  $\Wb_{vh}(\sigma_j - \sigma_k)$. (Replace $h$ by $p$ 
 or $q$ for the broken left and right half-lines.) Then the partition 
 function is
 \be \label{defZ} 
 Z \eq \sum_{\sigma} \, \prod W_{vh}(\sigma_i - \sigma_j)  \prod 
 \Wb_{vh}(\sigma_j - \sigma_k) \comma \ee
 the products being over all edges of each type, and the sum over all 
 $N^M$ values of the $M$ spins. We expect the partition function 
 per site
 \bd \kappa \eq Z^{1/M} \ed
 to tend to a unique limit as the lattice becomes large in both 
 directions.

 Let $a$ be a spin on a site near the centre of the lattice, as in 
 the figure,  and  $r$ be any integer.
 Then the thermodynamic average  of   $\omega^{r a}$ is 
 \be \label{avfa}
 \tilde{F}_{pq}(r) \eq \langle \omega^{r a} \rangle \eq Z^{-1} \, 
 \sum_{\sigma} \, \omega^{r a}
 \prod W_{vh}(\sigma_i - \sigma_j)  \prod  
 \Wb_{vh}(\sigma_j - \sigma_k) \period \ee
 We expect this to also tend to a limit as the lattice
 becomes large. 

 We could allow each vertical (horizontal) rapidity line $\alpha$ to 
 have a different rapidity $v_{\alpha}$ ($h_{\beta}$). If an edge 
 of $\cal L$ lies on lines with rapidities $v_{\alpha}$, $h_{\beta}$,
 then the 
 Boltzmann weight function of that edge 
 is to be taken as $W_{vh}(n)$ or $\Wb_{vh}(n)$, with
 $v = v_{\alpha}$ and $h = h_{\beta}$.

 The weight functions $W_{pq}(n)$,  $\Wb_{pq}(n)$ satisfy the star-
 triangle relation.\cite{BPAuY88} For this reason we are free to move
 the rapidity lines around in the plane, in particular to interchange 
 two vertical or two horizontal rapidity lines.\cite{RJB78} So long 
 as no rapidity line crosses the site  with spin $a$ while making 
 such rearrangements,  the  average $\langle  \omega^{r a} \rangle$ 
 is {\em unchanged} by the  
 rearrangement.\footnote{Subject to boundary conditions: here we are 
 primarily interested in
 the infinite lattice, where we expect the  boundary conditions
 to have no effect on the rearrangements 
 we consider.}

 All of the $v, h$ rapidity lines shown in Figure \ref{sqlattice} 
 are ``full'', in the sense that they extend without break from
 one boundary to another. We can move any such 
 line away from the central site
 to infinity, where we do not expect it to contribute to 
 $\langle  \omega^{r a} \rangle$. Hence in the infinite lattice limit
 $\tilde{F}_{pq}(r) = \langle  \omega^{r a} \rangle$  must be 
 {\em independent} of {\em all } the full-line $v$ and $h$
 rapidities.


 The horizontal rapidity line immediately below $a$ has different
 rapidity variables $p$, $q$ on the left and the right of the break below
 $a$. This means that we cannot use the star-triangle relation to move 
 it away from $a$.

 It follows that $\tilde{F}_{pq}(r)$ will in general depend on 
 $p$ and $q$, as well as on the `` universal'' constants $k$ or $k'$. 
 We are particularly interested in the case when $q = p$. Then
 the $p,q$ line is not broken, it can be removed to infinity, so
 \be \label{defMr}
 {\cal M}_r \eq \tilde{F}_{pp}(r) \eq  \langle  \omega^{r a} \rangle \eq 
 {\rm independent \; \; of }\; \; p  \period \ee

  These are the desired order parameters  of the chiral Potts model,
 studied by Albertini {\it et al}. By using this  ``broken rapidity line''
 approach, I was finally ably to verify their 
 conjecture (\ref{conj}) in 2005\cite{RJB05a,RJB05b}. 
 Here I shall  present some of the observations that enabled me to 
 do this.

 \subsection*{Automorphisms}

 
 There are various automorphisms that change $x_p, y_p \mu_p, t_p$ 
 while leaving the relations (\ref{xymu} ) still satisfied. Four that 
 we shall use are $R, S, M, V$, defined by:
 \ba \label{autos}
 \{x_{Rp}, y_{Rp}, \mu_{Rp}, t_{Rp} \} & = & \{ y_p,\omega x_p, 1/\mu_p, 
  \omega t_p \} \comma \nonumber \\
 \{x_{Sp}, y_{Sp}, \mu_{Sp}, t_{Sp} \} & = & \{ 1/y_p, 1/x_p, 
 \omega^{-1/2} y_p /(x_p \mu_p), 1/t_p \} 
  \comma \\
 \{x_{Mp}, y_{Mp}, \mu_{Mp}, t_{Mp} \} & = & \{ x_p, y_p, \omega \mu_p, 
  t_p \} \comma \nonumber \\
 \{x_{Vp}, y_{Vp}, \mu_{Vp}, t_{Vp} \} & = & \{ x_p, \omega y_p, \mu_p, 
  \omega t_p \} \period \nonumber \ea

 \subsection*{The central sheet $\cal D$ and its neighbours.}

 We shall find it natural, at least for the special case discussed 
 below, to regard $t_p$ as the independent variable, and 
 $x_p, y_p, \mu_p$ to be defined it terms of it by (\ref{xymu}).
 They are not single-valued functions of $t_p$: to make them 
 single-valued we must introduce $N$ branch cuts $B_0, B_1,
 \ldots, B_{N-1}$ in the complex
 $t_p$-plane
 as indicated in Figure (\ref{brcuts}). They are about 
 the points $1, \omega, \ldots, \omega^{N-1}$, respectively,

 \setlength{\unitlength}{1pt}
 \begin{figure}[hbt]
 \begin{picture}(420,260) (0,0)

 \put (50,125) {\line(1,0) {350}}
 \put (225,0) {\line(0,1) {250}}

 \put (325,125)  {\circle*{9}}
 \put (175,208)  {\circle*{9}}
 \put (175,42)  {\circle*{9}}

 \put (315,100)  {\Large 1}
 \put (185,214)  {\Large $\omega$}
 \put (184,32)  {\Large $\omega^2$}

 \put (358,100) {{\Large {$B_0$}}}
 \put (135,219) {{\Large {$B_1$}}}
 \put (134,22) {{\Large {$B_2$}}}

 \put (305,10) {{\Large {$t_p$-plane}}}
 \thicklines
 \put (295,124) {\line(1,0) {60}}
 \put (295,125) {\line(1,0) {60}}
 \put (295,126) {\line(1,0) {60}}

 \put (160,16) {\line(3,5) {30}}
 \put (160,17) {\line(3,5) {30}}
 \put (160,18) {\line(3,5) {30}}

 \put (160,234) {\line(3,-5) {30}}
 \put (160,233) {\line(3,-5) {30}}
 \put (160,232) {\line(3,-5) {30}}

 \thinlines

 \ \end{picture}
 \vspace{1.5cm}
 \caption{The cut $t_p$-plane for $N=3$.}
 \label{brcuts}
 \end{figure}
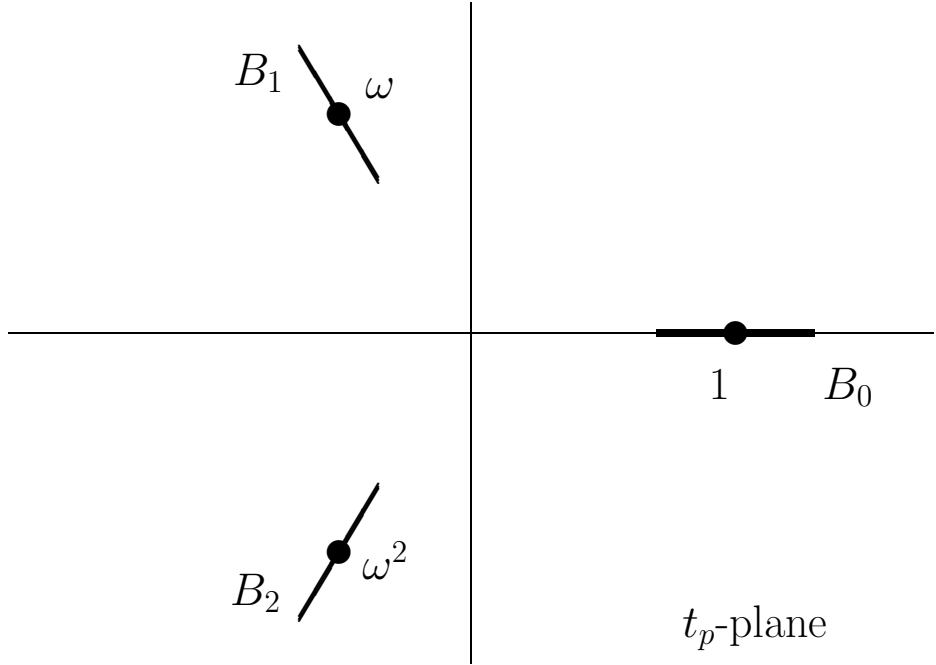

 Since the Boltzmann weights are rational functions of $x_p, y_p$, 
 we  expect $G_{pq}(r)$, considered as a function of $t_p$ or $t_q$, 
 to also have these $N$ branch cuts.

 Given $t_p$ in the cut plane of Figure \ref{brcuts}, choose
 $\mu_p^N$ to be outside the unit circle. Then $x_p$ must lie in 
 one of $N$ disjoint regions centred on the points
 $1, \omega, \ldots , \omega^{N-1}$. Choose it to be in the region
 centred on $1$. We then say that $p$ lies in the domain $\cal D$.
 When this is so (and $t_p$ is not close to a branch cut), then in 
 the limit $k' \rightarrow 0$,
 $\mu_p^N = O(1/k')$ and $x_p \rightarrow 1$.

 The domain $\cal D$ has $N$ neighbours ${\cal D}_0, \ldots,
 {\cal D}_{N-1}$ , corresponding to $t_p$ crossing the 
 $N$ branch cuts $B_0, \ldots, B_{N-1}$, respectively. The 
 automorphism that takes  $\cal D$ to  ${\cal D}_i$, while
 leaving $t_p$ unchanged, is
 \be  \label{defAi}
 A_i \eq  V^{i-1} R V^{N-i} \period \ee
 The mappings $A_i$ are involutions: $A_i^2 = 1$.




 \section{Functional relations}

 We define the ratio function
 \be \label{defGpq}
 G_{pq}(r) \eq \tilde{F}_{pq}(r) /\tilde{F}_{pq}(r-1)  \period \ee

 The functions $\tilde{F}_{pq}(r)$, $G_{pq}(r) $ satisfy two 
 reflection symmetry relations. Also, although we cannot move the 
 break in the $(p,q)$ rapidity line away from the spin $a$, we can
 rotate its parts about $a$ and then cross them over. As we show in 
 \cite{RJB98} and \cite{RJB05b}, this leads to functional relations 
 for $G_{pq}(r)$:
 \ba \label{functrl}
 G_{Rp,Rq}(r) & = & 1/G_{pq}(N-r+1) \comma \nonumber \\
 G_{p,q}(r) & = &  1/G_{RSq,RSp}(N-r+1) \comma  \nonumber \\
 G_{pq}(r)  & = &  G_{Rq, R^{-1} p}(r) \comma \\
 G_{pq}(r)  & = &  \frac{x_q \mu_q -  \omega^r x_p \mu_p }
 {y_p \mu_q -  \omega^{\, r-1} y_q  \mu_p} \; G_{R^{-1}q, R p}(r)  
 \nonumber \\
 G_{Mp,q}(r) & = & G_{p,M^{-1} q}(r)   =  G_{pq}(r+1) \comma 
 \nonumber \\
 \prod_{r=1}^N G_{pq}(r)  & = & 1 \period \nonumber \ea
 Also, {from} (\ref{defMr}),
 \be \label{calcMr}
 {\cal M}_r \eq G_{pp}(1) \cdots G_{pp}(r) \period \ee

 For the case when $N=2$ we regain the Ising model. As is shown in
 \cite{RJB98}, there
 is then a uniformizing substitution such that $x_p, y_p, \mu_p, t_p$
 are all single-valued meromorphic functions of a variable $u_p$, and
 $W_{pq}(n), \Wb_{pq}(n)$ and hence $G_{pq}(r)$ depend on $u_p$, 
 $u_q$ only via their
 difference $u_q - u_p$. In fact all quantities are Jacobi elliptic 
 functions of $u_p, u_q$ with modulus $k$. One can argue (based on
 low-temperature series expansions) that $G_{pq}(r)$ is analytic
 and non-zero in a particular vertical strip in the complex 
 $u_q - u_p$ plane. The relations (\ref{functrl}) then define 
 $G_{pq}(r)$. They can be solved by Fourier transforms and one readily 
 obtains the famous Onsager result
 \be {\cal M}_1 \eq (1-{k'}^2)^{1/8} \period \ee

 For $N > $ the problem is much more difficult. There then appears to
 be no uniformizing substitution and $G_{pq}(r)$ lives on a 
 many-sheeted Riemann surface obtainable from $\cal D$ by repeated 
 crossings of the branch cuts. One can argue from the physical cases
 (when the Boltzmann weights are real and positive) that $G_{pq}(r)$
 should be analytic and non-zero when $p, q$ both lie in 
 $\cal D$, but the  relations (\ref{functrl}) only relate these sheets 
 to a small sub-set of all possible sheets. There seems to be a basic 
 lack of information.




 \section{Solvable special case: $q = V p$}

 The author spent much time mulling over this problem, then towards the
 end of 2004 he realised that the case
 \be \label{spcase}
  q \eq Vp \ee
 may be much simpler to handle, and still be sufficient to obtain
 the order parameters ${\cal M}_r$.

 The reason it is simpler is that one can rotate the left-half line $p$ 
 anti-clockwise below $a$ until it lies immediately below the half-line
 $q$, as in Fig. 5 of \cite{RJB05b}. One has to reverse the direction 
 of the arrow, which means the rapidity is not $p$ but $p' = R^{-1}p$.


 \setlength{\unitlength}{1pt}
 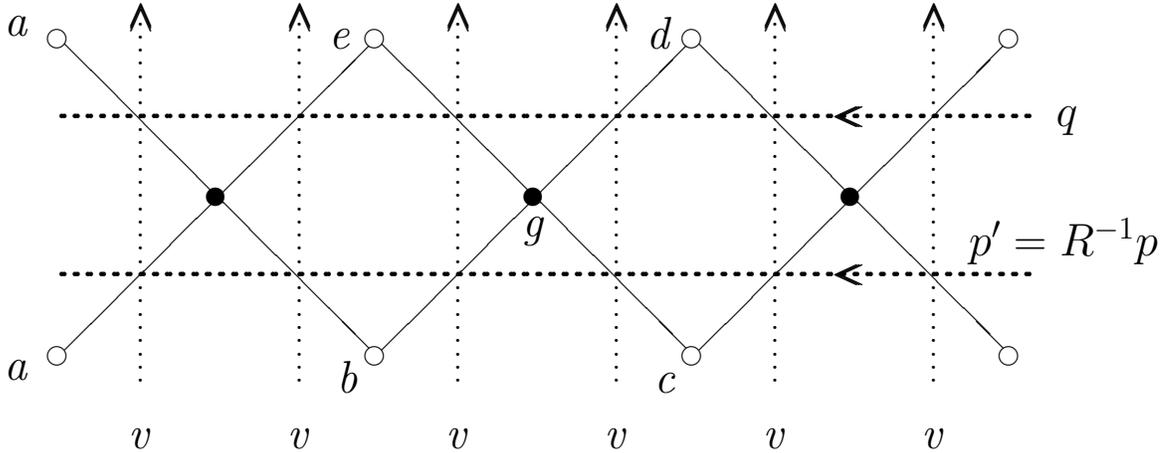
\begin{figure}[hbt]

 \begin{picture}(420,260) (0,0)

 \multiput(15,75)(5,0){74}{\bf .}
 \multiput(16,75)(5,0){74}{\bf .}

 \multiput(15,135)(5,0){74}{\bf .}
 \multiput(16,135)(5,0){74}{\bf .}

 \thicklines

 \put (308,72) {\large $< $}
 \put (309,72) {\large $< $}
 \put (310,72) {\large $< $}

 \put (308,132) {\large $< $}
 \put (309,132) {\large $< $}
 \put (310,132) {\large $< $}

 \put (42,170) {\large $\wedge$}
 \put (42,169) {\large $\wedge$}
 \put (42,168) {\large $\wedge$}

 \put (102,170) {\large $\wedge$}
 \put (102,169) {\large $\wedge$}
 \put (102,168) {\large $\wedge$}

 \put (162,170) {\large $\wedge$}
 \put (162,169) {\large $\wedge$}
 \put (162,168) {\large $\wedge$}

 \put (222,170) {\large $\wedge$}
 \put (222,169) {\large $\wedge$}
 \put (222,168) {\large $\wedge$}

 \put (282,170) {\large $\wedge$}
 \put (282,169) {\large $\wedge$}
 \put (282,168) {\large $\wedge$}

 \put (342,170) {\large $\wedge$}
 \put (342,169) {\large $\wedge$}
 \put (342,168) {\large $\wedge$}

 \thinlines


 \put (-5,166) {{\Large \it a}}
 \put (-5,36) {{\Large \it a}}
 \put (360,83) {{\Large $p' = R^{-1} p$}}
 \put (393,132) {{\Large $q$ }}

 \put (121,31) {{\Large \it b}}
 \put (241,31) {{\Large \it c}}
 \put (118,161) {{\Large \it e}}
 \put (238,161) {{\Large \it d}}
 \put (191,90) {{\Large \it g}}

 \put (195,105) {\circle*{7}}

 \put (18,163) {\line(1,-1) {115}}
 \put (138,163) {\line(1,-1) {115}}
 \put (258,163) {\line(1,-1) {115}}
 \put (18,47) {\line(1,1) {115}}
 \put (138,47) {\line(1,1) {115}}
 \put (258,47) {\line(1,1) {115}}

 \put (75,105) {\circle*{7}}
 \put (315,105) {\circle*{7}}

 \put (15,45) {\circle{7}}
 \put (135,45) {\circle{7}}
 \put (255,45) {\circle{7}}
 \put (375,45) {\circle{7}}

 \put (15,165) {\circle{7}}
 \put (135,165) {\circle{7}}
 \put (255,165) {\circle{7}}
 \put (375,165) {\circle{7}}

 \put (42,10) {{\Large \it v}}
 \put (102,10) {{\Large \it v}}
 \put (162,10) {{\Large \it v}}
 \put (222,10) {{\Large \it v}}
 \put (282,10) {{\Large \it v}}
 \put (342,10) {{\Large \it v}}

 \multiput(45,35)(0,5){28}{.}
 \multiput(105,35)(0,5){28}{.}
 \multiput(165,35)(0,5){28}{.}
 \multiput(225,35)(0,5){28}{.}
 \multiput(285,35)(0,5){28}{.}
 \multiput(345,35)(0,5){28}{.}
 \end{picture}

 \vspace{1.5cm}
 \caption{\footnotesize  The lattice after rotating 
 the half-line $p$ to a position  immediately below $q$.}
 \label{dblerow}
 \end{figure}

 The result is that $p$ enters the sums in (\ref{defZ}), 
	(\ref{avfa}) only via the weights of the edges shown in Figure 
	\ref{dblerow}. The left-hand spins are the same - the spin $a$. 
	The right-hand spins are set to the boundary value of zero.

	Further, we can sum over the spins between lines $p'$ and $q$. For 
	instance, summing over the spin $g$ gives a contribution
	\bd U(b,c,d,e) \eq \sum_g  W_{v p'} (b-g)   \Wb_{v p'} (c-g) 
	W_{v q} (g-d)   \Wb_{v q} (g-e)  \period \ed
	If $a, \sigma_1, \ldots , \sigma_L$ are the spins on the lowest row
	of Figure \ref{dblerow}, and $a, \sigma'_1, \ldots , \sigma'_L$ 
	are those in the upper, then the combined weight of the edges
	shown in Figure  \ref{dblerow} is
	\be \label{rowprod}
 \prod_{i=1}^L U(\sigma_{i-1},\sigma_i, \sigma'_i,\sigma'_{i-1})
	\period \ee

	Now
	$q = VRp'$, which from (\ref{autos}) means that
	\be x_q = y_{p'} \sep y_q = \omega^2 x_{p'} \sep \mu_q = 1/\mu_{p'} 
	\period \ee
	This is the equation (3.13) of \cite{BBP90}, the $q,r$ therein being
	our $p', q$  and $k, \ell$ having the values $0, 2$. From (3.17)
	therein, $U(b,c,d,e)$ vanishes if $0 \leq {\rm mod}(b-e,N) \leq 1$ and 
 $2 \leq {\rm mod}(c-d,N) \leq N-1$. It follows that the spins in the 
	upper  row are either equal to the corresponding spins in 
 the lower row,
	or just one less than them. From (2.29) and (3.39) of \cite{BBP90}, it 
	follows that to within ``gauge factors'' (i.e. factors that cancel out 
 of eqn. \ref{rowprod}) $U(b,c,d,e)$ depends on $p$ very simply: it is 
	{\em linear} in $t_p$.

	In fact, these Boltzmann weights $U(b,c,d,e)$ are those of the 
	$\tau_2(t_{p'})$ model\cite{BBP90,RJB90,RJB91} mentioned earlier.
	Just as this model plays a central role in the calculation of the 
	chiral Potts free energy, so it naturally enters this calculation of 
 the order parameters.

	In the low-temperature limit, when $k' \rightarrow 0$, 
 $\mu_p, \mu_q \sim
	O({k'}^{-1/N})$, $x_p, x_q \rightarrow  1$, we can verify that the 
	dominant contribution to the sums in (\ref{defZ}), 
	(\ref{avfa}) comes from the case when $\sigma_1, \ldots, \sigma_{L}, 
	\sigma'_1, \ldots, \sigma'_{L} $ are all zero. Also, to within
	factors that cancel out of(\ref{rowprod}) and  (\ref{avfa}),
	\be U(b,c,c,b)   = 1 - \omega t_{p'} = 1 - t_p \period \ee

	It follows that the RHS of (\ref{avfa}), and therefore of 
	(\ref{defGpq}), is a ratio of two polynomials 
	in $t_p$, each of degree $L$, and each equal to $(1-t_p)^L$ in the 
	limit $k' \rightarrow 0$. By continuity (keeping $L$ finite), for 
	small values of $k'$ their $L$ zeros must be  close to one. 
	Provided this remains true (which we believe it does)  when we take 
	the limit $L \rightarrow \infty$, we expect $G_{p,Vp}(r)$ to be
	an analytic and non-zero function of $t_p$, except in some region
	near $t_p = 1$. As $k'$ becomes small, this region must shrink
	down to the point $t_p = 1$. 

 Similarly, if  we rotate the half line $p$ in Figure 
 \ref{sqlattice} clockwise above $a$, we can move it be immediately
 above $q$, with $p$ replaced by $Rp$, as in Fig. 6 of \cite{RJB05b}.
 The $p', q$ of Figure\ref{dblerow} herein are now replaced by 
 $q, Rp$. This corresponds equation (3.13) of \cite{BBP90} with the 
 $q,r$ therein replaced by  $q, Rp$. From (\ref{spcase}) it follows 
 that $k, \ell$ in  \cite{BBP90} now have the values $-1, N+1$. The 
 combined star weights $U$ are now those of the $\tau_{N}(t_p)$
 model. They are polynomials in $t_p$ of degree $N-1$, except for  
 terms  which contribute a factor $x_p^{\epsilon(r)}$ to the 
 contribution of (\ref{rowprod}) to  $G_{p,Vp}(r)$, where
 \be \epsilon(r) \eq 1 - N \delta_{r,0} \comma \ee
 the $\delta$ function being interpreted modulo $N$, so 
 $\epsilon(0)  = \epsilon(N) = 1-N$. 

 When 
 $k' \rightarrow 0$ these polynomials are $(1-\omega t_p) 
 (1-\omega^2 t_p)  \cdots
 (1-\omega^{N-1} t_p)$. In the large-$L$ limit, with $k'$ not too 
 large, we therefore expect 
 $x_p^{\epsilon(r)} G_{p,Vp}(r)$ to have singularities near 
 $t_p = \omega, \ldots, 
 \omega^{N-1}$, but {\em not} near $t_p = 1$.

 If we define 
 \be  \label{greln}
 g(p;r) \eq G_{p,Vp}(r)  \comma \ee
 then this implies that the function  $ x_p^{\epsilon(r)} g(p;r)$
 does {\em not} have $B_0$ as a branch cut.
 This is in agreement with the fourth and sixth functional relations 
 in 
 (\ref{functrl}). If we set $q = Vp$ therein  we obtain
 \be \label{frln4}
 x_p^{-\epsilon(r)} g(p;r) \eq   y_p^{-\epsilon(r)} 
 g(V^{-1} R p;r) \comma \ee
 using $V^{-1}R = R^{-1} V$. Here we have used the fourth relation
 for $r \neq 0$ and the sixth to then determine the behaviour
 for $r=0$. (For $r=0$ the fourth relation merely gives $0 = 0$.)
 {From} (\ref{defAi}) the automorphism $V^{-1}R$ is the 
 automorphism $A_0$ that takes $p$ across the branch cut $B_0$,
 returning $t_p$ to its original value, while interchanging $x_p$ 
 with $y_p$. Thus (\ref{frln4}) states that 
 $ x_p^{-\epsilon(r)} g(p;r)$ is the same on both sides of the cut, 
 i.e. it does not have the cut $B_0$.


	These are the key analyticity properties that we need to calculate 
	$g(p;r)$ and ${\cal M}_r$. We do this in \cite{RJB05b,RJB05a}, 
 but this meeting is in honour of Tony Guttmann, an expert in series 
 expansion methods, so it seems appropriate to here describe the 
 series expansion checks I made (for $N=3$) when I first began to 
 suspect these properties.




 \section{Consequences of this analyticity}

 The above observations imply that $g(p;r)$, considered as a function 
 of $t_p$, does {\em not} have the branch cuts of Figure \ref{brcuts}, 
 except for the branch cut on the positive real axis. 

 This means that $g(p;r)$ is unchanged by taking allowing $t_p$ to 
 cross any of the branch cuts $B_1, \ldots ,B_{N-1}$ and then 
 returning it to its original 
 value, i.e. it satisfies the $N-1$ symmetry relations:
 \be \label{autosA}
  g(p;r) \eq g(A_i \, p;r)   \; \; \; {\rm for } \; \;  i = 1, 
 \ldots ,N-1 \comma \ee
 $A_i$ being the automorphism (\ref{defAi}).
 
 For $N = 3$, this can be checked using the series expansions
 obtained in \cite{RJB98b}. We use the hyperelliptic parametrisation
 introduced in \cite{RJB90b,RJB93a,RJB93b}. We define parameters
 $x, z_p, w_p$ related to one another and to $t_p$ by
 \be \label{defx}
 (k'/k)^2 = 27 x \prod_{ n =1}^{\infty} 
 \left( \frac{1-x^{3n}}{1-x^n} \right)^{12} \period \ee
  \be \label{eq4.5}
 w = \prod_{n=1}^{\infty} \frac{(1-x^{2n-1} z/w) (1-x^{2n-1} w/z)
 (1-x^{6n-5} zw) (1-x^{6n-1} z^{-1} w^{-1})} 
 {(1-x^{2n-2} z/w) (1-x^{2n} w/z)
 (1-x^{6n-2} zw) (1-x^{6n-4} z^{-1} w^{-1})} \ee
 (writing $z_p, w_p$ here simply as $z, w$), and
 \be \label{eq27}
 t_p =  \omega \frac{f(\omega z_p)}{f(\omega^2 z_p)} 
 = \frac{f(-\omega /w_p)}{f(-\omega^2/w_p)} 
 = \omega^2 \frac{f(-\omega w_p/z_p)}{f(-\omega^2 w_p/z_p)} 
 \comma \ee
 where $f(z)$ is the function
 \be f(z) \eq \prod_{n=1}^{\infty} (1-x^{n-1}z ) (1-x^n/z) 
 \period \ee
 

 Note that $x$, like $k'$, is a constant (not a rapidity variable)
 and is small at low temperatures. We develop expansions in powers of 
 $x$. For $p$ in $\cal D$, 
 the parameters $z_p, w_p$ are of order unity, so to leading
 order $w_p = z_p +1$, $x_p = 1$, $y_p = 
 (\omega - \omega^2 z_p)/(1- \omega^2 z_p)$.

 The automorphisms $R,S, V$ transform $z_p, w_p$ to 
 \bd
 z_{R p} = x z_p \sep z_{Sp} = 1/(x z_p) \sep  z_{V p} = -1/w_p \ed
\be w_{R p} = z_p/w_p \sep w_{Sp} = 1/(x w_p)
 \sep w_{V p} = z_p/w_p \comma \ee
 so from (\ref{defAi}), if $p_i = A_i p$ then
 \ba
 z_{p_0} = -1/(x w_p) , & z_{p_1} = -x w_p/z_p  , &   
 z_{p_2} = z_p \nonumber \\
 w_{p_0} = -1/(x z_p)  , & w_{p_1} = w_p  , &
 w_{p_2} = x z_p/w_p \period \ea
 If we write $ g(p;r)$ more explicitly as $g(z_p,w_p;r)$, then
 the relations (\ref{autosA}) become
 \addtocounter{equation}{1}
 \setcounter{storeeqn}{\value{equation}}
 \setcounter{equation}{0}
 \renewcommand{\theequation}{\arabic{storeeqn}\alph{equation}}
 \ba \label{eq1}
 g(z_p,w_p;r) & = &  g(-x w_p/z_p,w_p;r)  \\
 \label{eq2} g(z_p,w_p;r) & = &  g(z_p,x z_p/w_p;r) \period \ea
 \setcounter{equation}{\value{storeeqn}}
 \renewcommand{\theequation}{\arabic{equation}}

 Using (\ref{defZ}), (\ref{avfa}), we can write (\ref{defGpq}) as
 \be
 G_{pq}(r) \eq \sum_{j=0}^2 \omega^{jr } F_{pq}(j) \left/
 \sum_{j=0}^2 \omega^{j(r-1) } F_{pq}(j) \right. \comma \ee
 where $F_{pq}(j)$ is the probability that spin $a$ has value $j$.


 We use the series expansions (39) - (52) of \cite{RJB98b} for 
 $F_{pq}(1)/F_{pq}(0)$ and  $F_{pq}(2)/F_{pq}(0)$ in terms of 
 \be \alpha = z_q/z_p \sep \beta = w_q/w_p \period \ee
 Since $q = Vp$, $ z_{q} = -1/w_p$, $ w_{q} = z_p/w_p$ and we find 
 from (39) of \cite{RJB98b} that $u = -\omega \, w_p/z_p$. (Choosing 
 the cube 
 root for $u$ to ensure that  $F_{pq}(i)/F_{pq}(0)$ is real when
 $y_p = y_q =  0$ which is when $z_p = \omega^2$, $w_p = -\omega$: 
 we then regain the physically interesting  $q = p$ 
 case of eqn. \ref{defMr}. )
  For $p, q$ in $\cal D$, the parameters $z_p,w_p,z_q,w_q, \alpha, 
 \beta$ are all of order unity, we can then use the expansion 
 (48) of  \cite{RJB98b} to obtain

 \bd F_{pq}(1)/F_{pq}(0) = \omega^2 \psi_1(z_p) \eq \omega^2 
    \psi_2(-w_p) \comma \ed
 \be \label{F12}
 F_{pq}(2)/F_{pq}(0) = 
 \omega \psi_2(z_p) \eq \omega \psi_1(-w_p)  \comma \ee
 where
 \bd  \psi_1(z) = - (z+1) x + (z+1)^3x^2/z -
    (z^3+6 z^2+ 16 z +16 +4 z^{-1}+z^{-2}) x^3  \ed
 \bd
 + (z^4+11 z^3+ 41 z^2+85 z +81+25 z^{-1} 
   + 7 z^{-2}+z^{-3})x^4 + O(x^5) \comma \ed
 and
 \bd \psi_2(z)  = z x - (2 z+1 +z^{-1}) x^2 -
    (z^2-  8 z -2 - 3 z^{-1}-z^{-2}) x^3  \ed
 \bd
 - ( 2 z^3 - 5z^2+31 z+6 +14 z^{-1} 
  + 5 z^{-2}+z^{-3}) x^4 + O(x^5) \period \ed

 The automorphism (\ref{eq1}) interchanges $\cal D$ with 
 ${\cal D}_1$. To leading order in $x$, the mid-point is when
 $z_p = \i \, x^{1/2}, w_p = 1$. This is on 
 the boundary of the domain $\cal D$, in which  the series
 (48) of  \cite{RJB98b} was obtained, so the series is not 
 necessarily convergent at this point. Nevertheless, if we 
 take $z_p = O(x^{1/2})$ in the above two series,
 we find the terms  originally of order $x^j$  become of 
 order not larger than $x^{(j+1)/2}$. Extrapolating, this suggests
 that the series do still converge at the midpoint, so we can use  
 them to check whether the symmetry is satisfied.

 The first check occurs at order $x^{3/2}$, where both series
 contain a term
 \bd \pm \, (x z_p - x^2 w_p/z_p)  \ed
 (using the fact that to leading order $w_p = 1$ at the midpoint).
 This is indeed symmetric under $z_p \rightarrow -x w_p/z_p$.
 If we subtract this term from the series (using the expansion of 
 $w_p$ in terms of $z_p$), we can then check the behaviour at order 
 $x^2$, and similarly then at order $x^{5/2}$. All three checks are
 satisfied by both series.

 The perceptive reader will remark that (\ref{F12}) allows us to
 work with $w_p$ instead of $z_p$. Since $w_p$ is unchanged by $A_1$,
 the symmetry appears obvious. Indeed it is, but only because a quite
 remarkable event occurred in deriving these series, namely the $z$
 series contains no powers of $z+1$ as denominators,  and the 
 $w$ series contains no powers of $w-1$. If one expands $w$ in terms 
 of  $z$ (or $z$ in terms of $w$), then one does find such terms. 
 It is their absence from (\ref{F12}) that makes the series obviously 
 convergent near $w = 1$ 
 or $z = -1$. I have presented the argument in terms of $z_p$ to make 
 it clear that one does indeed have three non-trivial checks on the 
 symmetry to the available order of the  series expansion.

 Similarly, (\ref{eq2}) interchanges $\cal D$ with 
 ${\cal D}_2$, with mid-point $z_p = -1, w_p = \i \, x^{1/2}$. If one 
 now works with $w_p$ as the variable, one can verify to the same 
 three orders the symmetry $w_p \rightarrow x z_p/w_p$.


 So our series provide no less than six checks on the symmetries
 (\ref{eq1}), (\ref{eq2}). When I first observed this, I could see 
 the resemblance to the properties of the free energy of the
 $\tau_2(t_q)$ model. One such property is that $\tau_2(t_q) 
 \tau_2(\omega t_q) \cdots \tau_2(\omega^{N-1}t_q)$ is a rational 
 function of $x_q^N$, so I looked at the series for
 \ba \label{defL}
 {\cal L}(p;r) & = & \prod_{j\, = 0}^{N-1} g(V^j \, p;r)  
 \nonumber \\
 & = & g(z_p,w_p;r) \,  g(-1/w_p,z_p/w_p;r) \, g(-w_p/z_p,-1/z_p;r) 
 \period \ea
 Choosing an arbitrary value for $z_p$ and working to 30 digits of 
 accuracy, I soon found that the series (known to order $x^4$) fitted 
 with the simple formulae
 \be \label{Lconj}
 {\cal L}(p;0) = 1/x_p^2 \sep {\cal L}(p;1) = k^{1/3} x_p 
 \sep {\cal L}(p;2) = k^{-1/3} x_p \period \ee

 All this strongly suggested that I was on the right track. It did 
 not take long to justify my 
 observations for general $N$. For instance, if  $g(p;r)$ only has 
 the branch cut $B_0$, and  $x_p^{-\epsilon(r)} g(p;r)$ does not have 
 that cut, then $x_p^{-\epsilon(r)} {\cal L}(p;r)$ does not have the 
 cut
 $B_0$. But this function is unchanged by $p \rightarrow Vp$, which 
 rotates the $t_p$ plane through an angle $2 \pi/N$. Hence it cannot
 have any of the cuts $B_0, B_1, \ldots, B_{N-1}$. We do not expect 
 any other singularities (e.g. poles) for $p$ in $\cal D$, so the 
 function is 
 analytic in the entire $t_p$ plane. It is bounded (the Boltzmann
 weights $W, \Wb$ remain finite and non-zero as $y_p \rightarrow 
 \infty$, the  ratio $\mu_p/y_p$ remaining finite), so from 
 Liouville's theorem it is a constant (independent of $p$ but 
 dependent on $r$).

 We can relate these constants to the desired order parameters
 ${\cal M}_r$ in two ways, and then use these relations to calculate 
 the ${\cal M}_r$. When $y_p = y_q = 0$ and $x_p = k^{1/N}$, our 
 special case $q = Vp$ intersects with physically interesting case 
 $q = p$, so from (\ref{defMr}),
 \be x_p^{-\epsilon(r)} {\cal L}(p;r) \eq k^{-\epsilon(r)/N} \, 
 ({\cal M}_r/ {\cal M}_{r-1})^N
 \period \ee

 When $y_p = y_q = \infty$ ($\mu_p/y_p$ remaining finite) and 
 $x_p = k^{-1/N}$ we find
 not $q = p$ but $q = M^{-1}p$, which is related to $q = p$
 by the fifth of the functional relations (\ref{functrl}), giving
 \be x_p^{-\epsilon(r)} {\cal L}(p;r) \eq k^{\epsilon(r)/N} \, 
  ({\cal M}_{r+1}/ {\cal M}_{r})^N
 \period \ee

 The left-hand sides of these last two equations, being constants,
 are the same in both equations. We can therefore equate the two 
 right-hand sides, for $r = 1, \ldots, N-1$. Using the fact that
 ${\cal M}_0 = {\cal M}_N = 1$, we can solve for ${\cal M}_1, \ldots,
 {\cal M}_{N-1}$ to obtain
 \be {\cal M}_r \eq k^{r(N-r)/N^2} \; \; {\rm for \; \;} r = 0, 
  \ldots , N \comma \ee
 which verifies the conjecture (\ref{conj}) of Albertini {\it et al} 
 \cite{AMPT89}. For $N=3$ these results do of course agree with my 
 original conjectures (\ref{Lconj}).

 In \cite{RJB05b} I also show that one can calculate $G_{P,Vp}(r) 
  = g(p;r)$ by a Wiener-Hopf factorization, giving
 \be \label{gS}
  g(p;r) \eq k^{(N+1-2r)/N^2} \, {\cal S}_p^{\, \epsilon (r)} \ee
 for $r = 1, \ldots , N$, where
 \be \label{defS}
  \log {\cal S}_p  \eq - \frac{2}{N^2} \log k   + 
 \frac {1}{2 N  \pi  } \, 
 \int_0^{2 \pi}   \frac{k' \e^{\i\theta}}{1-k' \e^{\i\theta}} \, 
 \log [\Delta(\theta) - t_p] \,  {\rm d}\theta  \comma \ee
 and
 \be \Delta( \theta ) \eq [(1-2k' \cos \theta + {k'}^2 )/k^2]^{1/N} 
 \period \ee
 (This function ${\cal S}_p$ should not be confused with the 
 automorphism  $S$ defined in (\ref{autos}). 

  
 As is implied by the above equations, ${\cal S}_p$ satisfies the 
 product relation
  \be {\cal S}_p {\cal S}_{Vp} \cdots {\cal S}_{V^{N-1} p} \eq 
  k^{-1/N} x_p \period \ee
 Also, if one sets $q = Vp$ in the second of the relations 
 (\ref{functrl}), uses the identity $R S = M V R S V$ and
 the fifth relation, one obtains 
 $ g(p;r) g(RSVp;N-r) = 1$, from which we can deduce the symmetry
 \be 
 {\cal S}_p \, {\cal S}_{RSVp} = k^{-2/N^2} \period \ee
 
 For $N=3$ the automorphism $p \rightarrow RSVp$ takes $z_p, w_p$ to 
 $-w_p, -z_p$, so this relation can then be written
 \be {\cal S}(z_p,w_p ) {\cal S}(-w_p,-z_p) \eq k^{-2/9} \period \ee




 \section{Another interesting case: $q = V^2 p$}

 We now have the solution for $G_{pq}(r)$ for $q=p$ and for $q= Vp$.
 This suggests looking at one more case: $q= V^2 p$, where
 $y_q = \omega^2 y_p$. Similarly to section 5, we set
 $g_2(p;r) = G_{pq}(r)$ and 
\bd L_2(p;r) = \prod_{j=0}^{N-1} g_2(V^j p;r) \period \ed

 For $N = 3$ we have used the series expansions of \cite{RJB98b} to
 obtain for this case
 \be F_{pq}(1) = \omega \phi(w_p) \sep  F_{pq}(2) = \omega^2 
 \phi(1/w_p)  \comma \ee
 where
 \bd \phi(w) = (w - 1)x - (2 w^2 - 2 w + 1)x^2/w +
  (2 w^3 + 6 w^2 - 6 w + 1) x^3/w \ed
 \be
  - (2 w^4 + 8 w^3 + 24 w^2 - 22 w + 5) x^4/w + O(x^5) \period \ee
 As in the previous case, the coefficients are Laurent polynomials
 in $w$. There is no sign of any singularity near $w_p=1$, 
 $t_p = \omega$ so this suggests that $G_{pq}(r)$, considered as a 
 function of $t_p$, does not have the branch cut $B_1$. 

 Indeed, this is  a consequence of the third functional relation
 (\ref{functrl}). Setting $q = V^2 p$ therein, we obtain
 \bd g_2(p;r) \eq  g_2(A_1 p;r) \comma \ed
 which tells us that $g_2(p;r)$ is unchanged by taking $t_p$ 
 across the branch cut $B_1$ and returning it to its original value.
 This means that the cut $B_1$ is unnecessary.
 However,  $g_2(p;r)$ does appear to have the other two  cuts
 $B_0$ and $B_2$.


 To the 
 available four terms in the series expansion we found
 \bd 
  L_2(p;1) \eq x_p^2 \comma \ed
 and 
 \be
 L_2(p;0) \eq k^{-1/3} x_p^{-1}  h(z_p,w_p)^3 \sep L_2(p;2) 
 \eq k^{1/3}  x_p^{-1}  h(z_p,w_p)^{-3} \comma \ee
 where 
 \bd h(z,w) = 1 + (x^2 - 6 x^3 + 35 x^4) (w/z^2 + zw - 
 z/w^2 + 3) \ed
 \be + \; x^4 (w^2/z^4 + z^2/w^4 + z^2 w^2 - 3) +O(x^5) 
 \period \ee

 The result for $ L_2(p;1)$ looks encouraging, and indeed to the four
 available terms in the series expansion we also find
 \be \label{g2p1}
 g_2(p;1) = k^{2/9} \, {\cal S}_p \, {\cal S}_{Vp} \period \ee

 The results for $ L_2(p;0)$ and $ L_2(p;2)$ are not so encouraging and
 I have failed to find any obvious result for these or for $g_2(p;0)$,
 $g_2(p;2)$. In \cite{RJB05b} I conjecture that for general $N$
 the functions $G_{p,V^i p}(r)$ have a simple form as a product of 
 $\cal S$ functions provided $i=0, \ldots, N-1$ and 
 $r = 1, \ldots, N-i$. For other values of $i, r$ they remain a puzzle.
 (Except when  $i=1$ and $r = N $: this case can be deduced from the 
 sixth relation  of eqn \ref{functrl}.)

 If (\ref{g2p1}) is correct, then we have some information on the 
 function $L_{pq}(r) $ of eqn. 56 of \cite{RJB98}. From this and the 
 first equation of (\ref{functrl}),
 \be
 L_{pq}(r) = G_{pq}(r) G_{Rq,Rp}(r) =  G_{pq}(r)/G_{qp}(N-r+1) 
 \period \ee
 Setting $q= Vp$ and using (\ref{greln}), we obtain
 \be
 L_{pq}(r) = g(p;r)/g_2(Vp;N-r+1) \period \ee
 Taking $r=0$, it follows from (\ref{gS}) and (\ref{g2p1}) that
 \be
  L_{pq}(0) = k^{-4/9} /({\cal S}_p^2 \, {\cal S}_{Vp} \, 
    {\cal S}_{V^2 p} ) = k^{-1/9}/(x_p{\cal S}_p) \period \ee 
 The function  $L_{pq}$, for arbitrary $p, q$, was introduced in  
 \cite{RJB98} partly because
 its square  is a rational function of 
  $x_p, y_p$, $\mu_p$, $x_q, y_q$, $\mu_q$ when $N=2$, so the hope was 
 that it might be  similarly simple for all $N$. We see that this 
 cannot be so: ${\cal S}_p$ is {\em not} such a function.





 \section{Summary}

 I have outlined the recent derivation of the order parameters
 of the solvable chiral Potts model, a derivation that verifies
 a long-standing and elegant conjecture.\cite{AMPT89} As with all the 
 calculations on solvable models satisfying the star-triangle 
 relations, the trick is to generalize the model to a point where
 one has a function, here $G_{pq}(r)$, to calculate, rather than a 
 constant, as one can obtain relations and properties that define 
 this function. On the other hand, this is an example where it
 pays {\em not} to over-generalize: we can handle the particular 
 function $G_{p,Vp}(r)$, and this is 
 sufficient for the purpose of obtaining the order parameters.
 The general $G_{pq}(r)$ continues to defy calculation.

  Series expansion methods can provide a valuable check 
 on such derivations, which are of their nature believable but hard 
 to make fully mathematically rigorous.
 One usually tries to present the argument in as logical a manner 
 as possible, but this is usually {\em not} the manner in which it was 
 originally developed. Here I have indicated the points in the 
 calculation when I found the available checks both reassuring and 
 encouraging.


 \smallskip

 \end{document}